\documentclass[twocolumn]{aastex62}
\usepackage{mathtools}
\usepackage{graphicx}
\usepackage{physics}
\usepackage{xcolor}
\usepackage{hyperref}

\received{2018 October 25}
\revised{2019 May 19}
\accepted{2019 June 11}
\published{2019 July 31}
\submitjournal{ApJ}

\shorttitle{Machine Learning Applied to the 21 cm Signal}
\shortauthors{La Plante \& Ntampaka}

\begin{document}

\title{Machine Learning Applied to the Reionization History of the Universe in
  the 21 cm Signal}

\correspondingauthor{Paul La Plante}
\email{plaplant@sas.upenn.edu}

\author[0000-0002-4693-0102]{Paul La Plante}
\affiliation{Center for Particle Cosmology \\
  Department of Physics and Astronomy \\
  University of Pennsylvania \\
  Philadelphia, PA 19104 USA}
\author[0000-0002-0144-387X]{Michelle Ntampaka}
\affiliation{Harvard-Smithsonian Center for Astrophysics\\
  Harvard University\\
  Cambridge, MA 02138, USA}
\affiliation{Harvard Data Science Initiative\\
  Harvard University\\
  Cambridge, MA 02138, USA}

\begin{abstract}
  The Epoch of Reionization (EoR) features a rich interplay between the first
  luminous sources and the low-density gas of the intergalactic medium (IGM),
  where photons from these sources ionize the IGM. There are currently few
  observational constraints on key observables related to the EoR, such as the
  midpoint and duration of reionization. Although upcoming observations of the
  21 cm power spectrum with next-generation radio interferometers such as the
  Hydrogen Epoch of Reionization Array (HERA) and the Square Kilometre Array
  (SKA) are expected to provide information about the midpoint of reionization
  readily, extracting the duration from the power spectrum alone is a more
  difficult proposition. As an alternative method for extracting information
  about reionization, we present an application of convolutional neural networks
  (CNNs) to images of reionization. These images are two-dimensional in the
  plane of the sky, and extracted at a series of redshift values to generate
  ``image cubes'' that are qualitatively similar to those HERA and the SKA will
  generate in the near future. Additionally, we include the impact that the
  bright foreground signal from the Milky Way galaxy imparts on such image cubes
  from interferometers, but do not include the noise induced from
  observations. We show that we are able to recover the duration of reionization
  $\Delta z$ to within 5\% using CNNs, assuming that the midpoint of
  reionization is already relatively well constrained. These results have
  exciting impacts for estimating $\tau$, the optical depth to the CMB, which
  can help constrain other cosmological parameters.
\end{abstract}


\section{Introduction}
\label{sec:intro}
The Epoch of Reionization (EoR) is a portion of the Universe's history
characterized by a large scale phase change of the intergalactic medium (IGM)
from neutral gas to ionized. The neutral hydrogen gas in the IGM emits radiation
at radio frequencies with a wavelength of $\lambda = 21$ cm due to the hyperfine
transition of the ground state, which is being pursued observationally through
radio interferometer telescopes such as the Hydrogen Epoch of Reionization Array
(HERA\footnote{\url{https://reionization.org}}), the Low Frequency Array
(LOFAR\footnote{\url{https://www.lofar.org}}), and the Square Kilometre Array
(SKA\footnote{\url{https://www.skatelescope.org}}). These arrays seek to
generate images of the EoR, which would provide insight into the topology of
reionization and yield vital clues as to the astrophysical sources responsible
for reionization. The images generated by these instruments can be useful for
learning valuable information about these sources; for instance, locating highly
ionized regions can pinpoint regions to perform follow-up analysis with other
instruments such as the James Webb Space Telescope, or for cross correlating
with data from other surveys.

These images from the EoR can also be used to help constrain key properties of
reionization, such as the midpoint and duration of reionization. However,
extracting these parameters directly from the images may be challenging, due to
the large degree of contamination from bright foreground emission from the Milky
Way galaxy. This emission is typically several orders of magnitude larger than
the target signal, making simple imaging of the sky impossible. Due to their
spectral smoothness, these foregrounds are mostly constrained to low
$k_\parallel$-values in Fourier space. One approach for extracting information
about reionization from such observations is to compute the power spectrum using
Fourier modes that are uncontaminated by the foregrounds. Importantly, removing
the $k_\parallel=0$ mode precludes referencing an absolute scale, and so the
resulting images are fluctuations about the mean temperature of the map rather
than an absolute scale. However, such approaches do not leverage all of the
information present in the images produced by these instruments. In particular,
the power spectrum is insensitive to any non-Gaussian information present in the
images. The 21 cm field is expected to be highly non-Gaussian during the EoR,
and so there is valuable information that such approaches are potentially
insensitive to.

An alternative approach to computing power spectra is to apply machine learning
techniques of image processing to simulated maps of the EoR. Such an approach
allows for extracting non-Gaussian information present in the maps, without
having to resort to explicitly computing higher-point statistics. We present
here the use of convolutional neural networks (CNNs) to extract key features of
the reionization history in semi-numeric realizations of the EoR. In particular,
we use CNNs to regress on the parameters of the midpoint of reionization
$z_\mathrm{50}$ and duration of reionization $\Delta z$. These two parameters
are sufficient to characterize many of the key features of the reionization
history, such as the optical depth to the CMB $\tau$. Constraining $\tau$ is
useful in a broader cosmological context: providing tight priors on $\tau$ from
observations of the EoR can yield smaller uncertainties on other cosmological
parameters, specifically $A_s$, when analyzing CMB data \citep{liu_etal2016a}.

Machine learning has been used in a variety of cosmological and astrophysical
applications, including utilizing weak lensing beyond 2-point statistics to
produce tighter cosmological parameter constraints \citep{gupta_etal2018},
parameter inference from weak lensing maps \citep{ribli_etal2018}, classifying
reionization sources \citep{hassan_etal2018}, identify lensing signals in images
\citep{lanusse_etal2018}, reducing errors in cluster dynamical mass measurements
\citep{ntampaka_etal2015}, and morphological classification of galaxies
\citep{dieleman_etal2015}.  Convolutional Neural Networks \citep[CNNs,
e.g.~][]{fukushima_miyake1982,lecun_etal1999,krizhevsky_etal2012} are a class of
machine learning algorithms that are typically used for image recognition and
classification tasks.  These networks are typically deep, utilizing many hidden
layers to extract features from the input images by learning filters, weights,
and biases to minimize a loss function on labeled training data. The authors of
\citet{gillet_etal2018} have used CNNs to extract semi-analytic reionization
model parameters. This work builds on their findings, and demonstrates that
information about the reionization history can be extracted even when accounting
for some of the effects from observational foregrounds.

We organize the paper as follows. In Sec.~\ref{sec:images}, we describe the
reionization model used and the method by which the input images are
generated. In Sec.~\ref{sec:ml_methods}, we describe the machine learning
approach we employed. In Sec.~\ref{sec:results}, we present the results of our
trained CNN and its ability to reconstruct the reionization history. In
Sec.~\ref{sec:discussion}, we provide discussion of interpreting the
intermediate output of CNNs. In Sec.~\ref{sec:conclusion}, we
conclude. Throughout the work, we employ a $\Lambda$CDM cosmology with
cosmological constants given by the \textit{Planck} 2015 results
\citep{planck2015}.

\section{Making Images of the EoR}
\label{sec:images}
The input to the CNNs discussed later are typically two-dimensional ``images'',
to which various mathematical functions such as convolutions are
applied. Because the 21 cm field in principle provides a tomographic, fully
three-dimensional picture of the Universe as a function of wavelength/redshift,
the input images must be converted to be two-dimensional by some method. In
\citet{gillet_etal2018}, the authors take one dimension to be in the plane of
the sky, and the other to be along the line-of-sight axis. This approach
generates images of reionization ``from the side,'' in a sense, where the light
cone-induced redshift evolution is explicitly presented to the CNN as one of the
axes in the input images. However, actual images from interferometers in the
near future will suffer from effects related to foreground contamination
\citep{beardsley_etal2015}, and so this work endeavors to include a subset of
such effects to investigate their impact on the ability of CNNs to extract
useful information.

The approach presented in this work is to leverage the fact that CNNs can take
multiple two-dimensional images as input. Each two-dimensional image is referred
to as a ``channel,'' due to their typical usage of providing color information
to CNNs in traditional image-recognition applications. As we outline in detail
below, we generate a two-dimensional image of 21cm signal in the plane of the
sky, taking into account several foreground features imparted by
interferometers. We generate images at multiple different redshifts, and pass
each redshift image into the CNN as a separate channel. This approach results in
input images that are more representative of what current and next-generation
interferometers will be capable of than have been presented previously in the
literature.

To provide input images to the CNNs discussed later, suitable images must first
be generated representing reionization. To explore the effects of foreground
contamination, we include several key instrumental effects, such as a finite
angular resolution specific to HERA and the foreground wedge. We begin by
discussing the semi-numeric reionization model used in this project, then the
observational effects included in the images, and finally how realizations were
generated.

\subsection{Reionization Model}
\label{sec:zreion}
The reionization model used in this work is based off of
\citet{battaglia_etal2013a}. It has been used to model the EoR
\citep{laplante_etal2014}, and to understand the impact of patchy reionization
on the kSZ effect \citep{battaglia_etal2013b,natarajan_etal2013}. We summarize
here some of the key features of the model, and refer to the other papers for a
detailed discussion of the method.

The model of \citet{battaglia_etal2013a} assumes the ``redshift of reionization
field'' $z_\mathrm{re}(\vb{x})$, defined as
\begin{equation}
  \delta_z(\vb{x}) \equiv \frac{\qty[z_\mathrm{re}(\vb{x}) + 1] - \qty[\bar{z} + 1]}{\bar{z} + 1},
  \label{eqn:deltaz}
\end{equation}
is a biased tracer of the dark matter overdensity field
\begin{equation}
  \delta_m(\vb{x}) \equiv \frac{\rho_m(\vb{x}) - \bar{\rho}_m}{\bar{\rho}_m}
  \label{eqn:deltam}
\end{equation}
on large scales ($\geq 1$ $h^{-1}$Mpc). To quantify the precise relationship
between the fields, a bias parameter $b_{zm}(k)$ is introduced:
\begin{equation}
  b_{zm}^2(k) \equiv \frac{\ev{\delta^*_z \delta_z}_k}{\ev{\delta^*_m \delta_m}_k} = \frac{P_{zz}(k)}{P_{mm}(k)}.
\end{equation}
We parameterize the bias parameter $b_{zm}(k)$ as a function of spherical
wavenumber $k$ in the following way:
\begin{equation}
  b_{zm} = \frac{b_0}{\qty(1 + \frac{k}{k_0})^\alpha}.
  \label{eqn:bias}
\end{equation}
The value $b_0$ can be predicted using excursion set formalism in the limit
$k \to 0$ \citep{barkana_loeb2004}. We use the value of
$b_0 = 1/\delta_c = 0.593$. The reionization field for a given density field is
then completely specified by the three values of the parameters $\bar{z}$ in
Eqn.~(\ref{eqn:deltaz}), which determines the midpoint of reionization, and
$k_0$ and $\alpha$ in Eqn.~(\ref{eqn:bias}), which determine the duration.

An important feature to point out of the semi-analytic model is that the
redshift when the volume is 50\% ionized $z_{50}$ is in general not exactly
equal to $\bar{z}$ in Eqn.~(\ref{eqn:deltaz}). This is due to the fact that the
distribution of redshift values in the $z_\mathrm{re}(\vb{x})$ field are not
symmetric about the mean, and so the median redshift is not equal to the mean.
As such, throughout the rest of this paper, we will refer to the semi-analytic
model parameter as $\bar{z}$ and the redshift when the volume is 50\% ionized
(by volume) as $z_{50}$. Additionally, to quantify the duration of reionization
$\Delta z$, we compute the difference in redshift between when the volume is
25\% ionized $z_{25}$ and 75\% ionized $z_{75}$. Mathematically,
\begin{equation}
  \Delta z \equiv z_{25} - z_{75}.
\end{equation}
Other works have introduced parameters to quantify the asymmetry of reionization
\citep{trac2018}, but for the current work we deal only with the midpoint and
duration. These two numbers capture many of the physical reionization scenarios
precipitated by Population~II stars. So-called ``exotic'' reionization scenarios
featuring Population~III stars or mini-quasars are likely not accurately
captured by this relatively simple semi-numeric model of reionization, but we
leave an investigation of these scenarios to future work.

\subsection{Observational Effects}
\label{sec:foregrounds}
Map-making of the EoR using interferometers such as HERA or the SKA will include
several interesting observational effects. First, the angular resolution of the
instrument, determined by the longest baseline in the array, sets a lower limit
on the spatial scales that can be probed by the instrument. Given a baseline
vector connecting two interferometer elements $\vb{b}$, associated coordinates
in the $uv$-plane are \citep{thompson_etal2001}:
\begin{equation}
  \vb{u} = \frac{\vb{b}}{\lambda},
\end{equation}
where $\lambda$ is the wavelength of the radiation of interest. For the
redshifted 21 cm line, this is simply $\lambda = \lambda_0 (1+z)$, with
$\lambda_0 = 21$ cm. The $uv$-coordinates are related to the comoving wavenumber
$k_\perp$ in the plane of the sky through \citep{thyagarajan_etal2015}:
\begin{equation}
  \vb{k}_\perp = \frac{2\pi \vb{u}}{D_c},
  \label{eqn:kperp_max}
\end{equation}
where $D_c$ is the comoving distance along the line of sight to the observed
redshift:
\begin{equation}
  D_c(z) = c \int_0^z \frac{\dd{z'}}{H(z')},
\end{equation}
where $H(z)$ is the Hubble parameter. In general, an interferometer is not
sensitive to information for Fourier modes with
$k_\perp \geq k_{\perp,\mathrm{max}}$ defined by its largest baseline. For the
fully constructed HERA-350 array, the longest baseline will be $\sim$870 m
\citep{deboer_etal2017}. This corresponds to
$k_{\perp,\mathrm{max}} \sim 0.5\ h^{-1}$ Mpc at $z = 8$. To generate images
that reflect this limited angular resolution, we apply a mask in Fourier space
where the value of all modes is set to 0 where
$k_\perp > k_{\perp,\mathrm{max}}$ (discussed below in detail in
Sec.~\ref{sec:image_method}).

Another more subtle observational effect is the impact that the aforementioned
smooth foregrounds has on the signal. The dominant signal in the radio portion
of the electromagnetic spectrum at these wavelengths is ``foreground'' emission
from our own Milky Way galaxy. Specifically, the galactic synchrotron radiation
has a brightness temperature of several thousand Kelvin at 150 MHz. Because
synchrotron radiation follows a power law as a function of frequency, it is very
smooth in Fourier space. Naively, all of the power from these foregrounds would
fall into bins of small values of $k_\parallel$.

Unfortunately, due to the chromaticity of an interferometer, the power from
these small-$k_\parallel$ modes scatters to higher values of $k_\parallel$. (For
an explanation of why this happens, see \citealt{parsons_etal2012}.) The amont
of contamination is a function of $k_\perp$, and increases sharply as $k_\perp$
increases. The resulting foreground contamination is affectionately referred to
as ``the wedge'' in the literature
\citep{datta_etal2010,vedantham_etal2012,morales_etal2012,liu_etal2014a}. The
slope of the wedge $m$ in $k_\perp$-$k_\parallel$ space is largely independent
of the specifics of the instrument, and can be written as
\citep{thyagarajan_etal2015}:
\begin{equation}
  m(z) \equiv \frac{k_\parallel}{k_\perp} = \frac{\lambda D_c f_{21} H}{c^2
    (1+z)^2},
  \label{eqn:wedge_slope}
\end{equation}
where $\lambda = \lambda_0 (1+z)$ is the wavelength of the 21 cm signal at the
redshift of interest, $D_c$ is the comoving distance to a redshift $z$, $f_{21}$
is the rest-frame frequency of the 21 cm signal, and $H$ is the Hubble parameter
of redshift $z$. For the redshifts of interest, $m \sim 3$. In this expression,
we have assumed the maximal horizon contamination: physically, the bright
foreground contamination extends down to the horizon of the interferometer
beam. \footnote{The foreground contamination actually extends slightly beyond
  the horizon limit. This ``supra-horizon buffer'' is due to the intrinsic
  spectral unsmoothness of the foreground signal \citep{pober_etal2013}. The
  amount of additional leakage is relatively small, especially for the longest
  baselines which are most important for forming images.} In practice, with
foreground mitigation and removal schemes, the slope of the wedge may not be as
steep as the expression in Eqn.~(\ref{eqn:wedge_slope}). If the contamination is
not as dire as that specified by Eqn.~(\ref{eqn:wedge_slope}), then the accuracy
of the predictions should only increase, because there is more information
available to the CNN in the input images during training. Conversely, if more
modes are contaminated than those which na\"\i vely should be foreground-free,
the resulting images of the EoR from interferometers such as HERA may in fact be
worse than those presented here. The quality of actual images will depend on the
processing details of mapmaking using the visibility data, and so future studies
may require a more nuanced treatment of foreground contamination.

\subsection{EoR Input Image Generation}
\label{sec:image_method}

\begin{figure*}[t]
  \centering
  \includegraphics[width=0.9\textwidth]{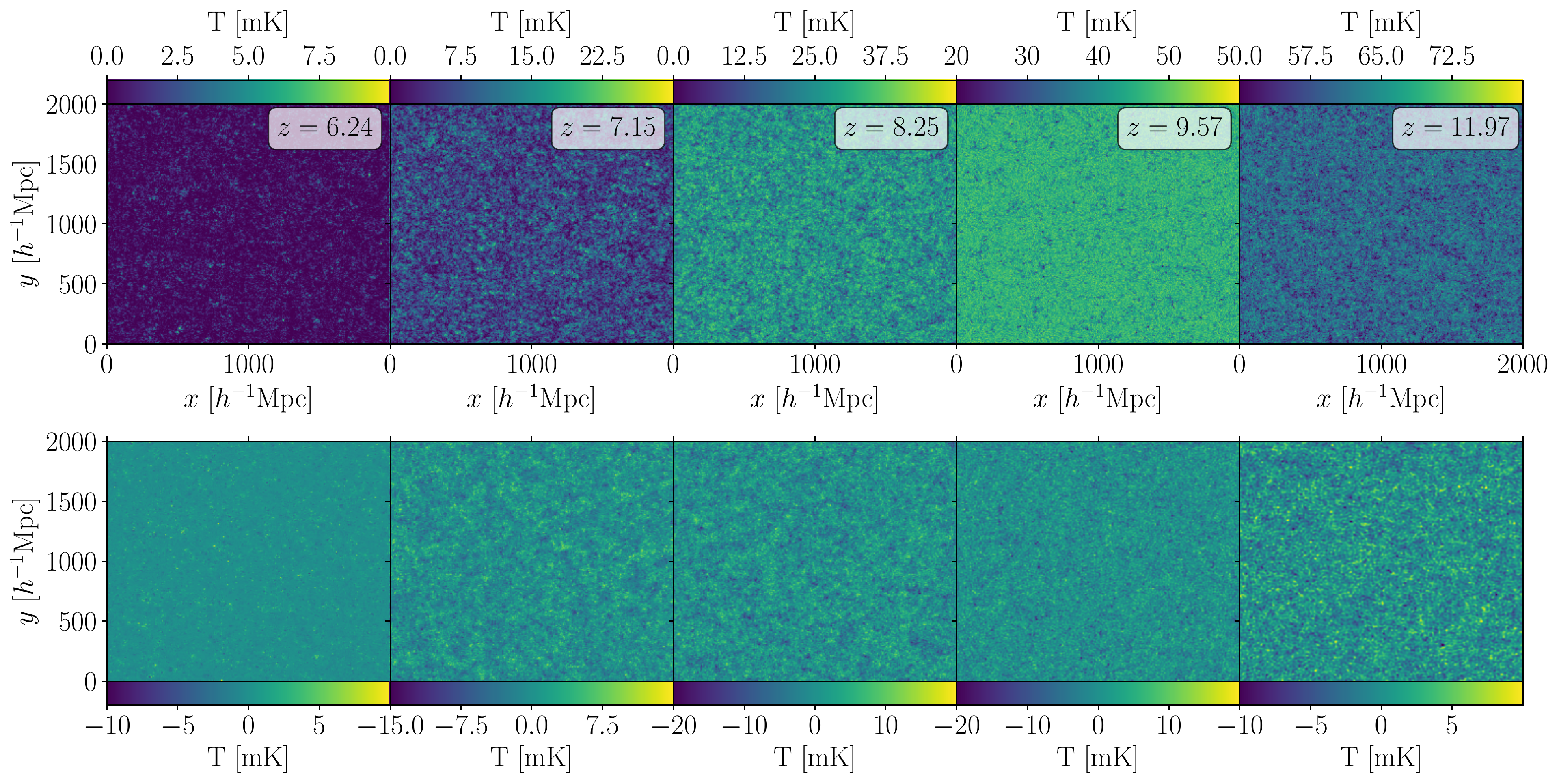}
  \caption{A visualization of the 21 cm images before (top) and after (bottom)
    the application of the foreground effects described in
    Sec.~\ref{sec:foregrounds}. The different columns are different redshift
    slices, with the redshift shown on the image. The most dramatic change is
    that the zero-level is no longer the absence of the 21 cm signal, as in the
    top row, but the mean value of the Fourier-transformed slab. This effect is
    due to the removal of the $k_\parallel=0$ mode as part of the foreground
    effects, which ensures that the resulting inverse-Fourier transformed slab
    must have mean 0. Also note that the structures in the top panel are no
    longer in the corresponding places in the bottom panel, another effect of
    the application of the foreground wedge. Although this effect makes matching
    the locations of individual sources difficult, statistically, the fields
    seem to have similar properties. See Sec.~\ref{sec:image_method} for more
    discussion.}
  \label{fig:viz_layers}
\end{figure*}

To actually construct the reionization realizations used as input images, we
first perform an $N$-body simulation to generate the dark matter density field
$\delta_m$ used in Eqn.~(\ref{eqn:deltam}). The $N$-body simulation uses a
P$^3$M algorithm described in \citet{trac_etal2015}, and contains 2048$^3$ dark
matter particles in a volume of 2 $h^{-1}$Gpc on a side. At regularly spaced
intervals in redshift, the particles are deposited on a grid using a
cloud-in-cell (CIC) scheme onto a uniform grid of 2048$^3$ resolution
elements. In order to obtain the density field at an arbitrary redshift, the two
neighboring matter density fields are loaded into memory, and interpolated in
scalefactor $a$ for every point in the volume. This allows for the construction
of an approximate density field for any desired redshift without having to run a
new simulation \textit{ab initio}.

To generate a new series of input images, a new set of model parameters
\{$\bar{z}$, $k_0$, $\alpha$\} are chosen. The density field $\delta_m(\vb{x})$
is generated for the mean redshift $\bar{z}$, according to the scheme described
in the previous paragraph. Then, the density field is Fourier transformed into
$k$-space to generate $\delta_m(\vb{k})$, and the bias relation defined in
Eqn.~(\ref{eqn:bias}) is applied as a function of spherical wavenumber $k$ to
generate $\delta_z(\vb{k})$.\footnote{As in \citet{battaglia_etal2013a}, the CIC
  window from particle deposition is deconvolved, and the field is smoothed
  using a spherical tophat window with a radius of 1 $h^{-1}$Mpc.} The field is
inverse Fourier transformed to arrive at $\delta_z(\vb{x})$, and
Eqn.~(\ref{eqn:deltaz}) is inverted to get the field
$z_\mathrm{re}(\vb{x})$. This field encodes the entire reionization history for
the volume given the density field and model parameters chosen.

\begin{figure*}
  \includegraphics[width=\textwidth]{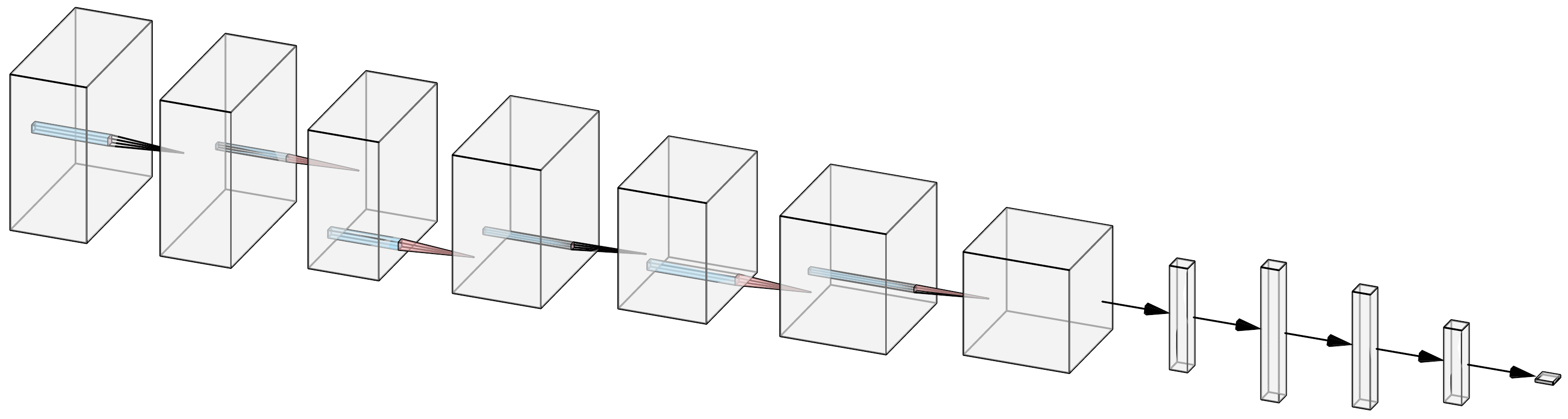}
  \caption{A visualization of the convolutional neural network (CNN)
    architecture used in the analysis. The input images are
    $512 \times 512 \times 20$, and the output is two meta-parameters $z_{50}$
    and $\Delta z$ quantifying the midpoint and duration of reionization,
    respectively. As shown in the figure, there are many hidden layers between
    the input and output. The addition of hidden layers has been shown to
    increase the efficacy of the network in processing image-like data
    \citep{schmidhuber2014}. See Sec.~\ref{sec:ml_methods} for a detailed
    description of the full architecture of the network.}
  \label{fig:architecture}
\end{figure*}

As discussed in Sec.~\ref{sec:images}, the next task is to generate the 21 cm
field $T_{21}$ at a fixed series of redshifts, and apply the foreground effects
described in Sec.~\ref{sec:foregrounds}. This process generates input images
which attempt to be representative of images generated by HERA in the next 2-3
years--images from a next-generation experiment, such as the SKA, will
presumably provide images with even higher sensitivity and fidelity. Given a
snapshot $i$ with redshift $z_i$, the matter density field is generated by
interpolating using bracketing redshifts as described earlier to generate
$\delta_m(\vb{x}, z_i)$. The 21 cm brightness temperature field $\delta T_b$
corresponding to this redshift is then generated by using the formula
\citep{madau_etal1997}: \begin{multline}
  \delta T_b = 26(1 + \delta_m)x_\mathrm{H\textsc{i}} \qty(\frac{T_S - T_\gamma}{T_S}) \qty(\frac{\Omega_b h^2}{0.022}) \\
  \times \qty[\qty(\frac{0.143}{\Omega_m h^2}) \qty(\frac{1+z}{10})]^{\frac{1}{2}} \, \mathrm{mK},
  \label{eqn:t21}
\end{multline}
where $x_\mathrm{H\textsc{i}}$ is the neutral fraction field for a given point
in the volume, $T_S$ is the spin temperature of the gas, and $T_\gamma$ is the
temperature of the CMB at redshift $z_i$. For simplicity, we assume that the
ionization field only takes on values of 0 (for a totally ionized part of the
volume) or 1 (for a neutral region). In particular, if
$z_\mathrm{re}(\vb{x}) < z_i$, then the model predicts a later redshift of
reionization than the redshift in question, and so that part of the volume is
neutral. Also, we assume that the spin temperature is much greater than the CMB
temperature, so $(T_S - T_\gamma)/T_S \to 1$. This assumption is valid once the
spin temperature is coupled to the kinetic temperature of the gas, which happens
once the gas is ionized at the $\sim$25\% level \citep{santos_etal2008}. As
shown in \citet{greig_mesinger2017}, incorrectly assuming spin temperature
saturation can bias the recovery of semi-analytic model parameters, especially
when applied at high redshift ($z \gtrsim 15$) when the spin temperature is not
saturated. Because we are interested in extracting the timing of the midpoint
and central duration of reionization, the assumption of spin temperature
saturation was used in the application at hand. It should be noted, though, that
this model is overly na\"{i}ve for understanding how the brightness temperature
behaves at the very onset of reionization, and may be inaccurate for the highest
redshift layers in our input images ($z \sim 12$). A more detailed treatment of
the spin temperature should be used if applying this technique to simulated
images of the pre-reionization epoch.

Once a new reionization history has been generated given the model parameters
$\{\bar{z}, \alpha, k_b\}$, a series of snapshot images is generated. To avoid
biasing the results of the CNN by tracing out the same density structure with
different reionization histories, starting indices $(i, j, k)$ are chosen
randomly, as well as a random line-of-sight along the $x$-, $y$-, or
$z$-axis. Starting at these coordinates, a series of 20 snapshots spanning the
range $5.5 \leq z \leq 12$ are chosen along the line of sight, evenly divided in
comoving distance. At a given redshift snapshot $z_i$, a two-dimensional slab is
generated, with the axes in the plane of the sky spanning the full size of the
simulation volume (2 $h^{-1}$Gpc $\times$ 2 $h^{-1}$Gpc) and the axis along
the line of sight spanning $\sim 48.8\ h^{-1}$Mpc. The redshift evolution along
this distance is small enough that the slab can be approximately considered
comoving, and so there is no light cone effect induced
\citep{laplante_etal2014}. This slab is then Fourier transformed, and
Equations~(\ref{eqn:kperp_max}) and (\ref{eqn:wedge_slope}) are applied (the
maximum angular resolution observable for HERA and the effect of foreground
contamination, respectively). Specifically, all Fourier modes where
$k_\perp > k_{\perp,\mathrm{max}}$ for HERA are removed, as well as all modes
for which $k_\parallel \leq m k_\perp$. Additionally, to reduce the final size
of the output images, we downselect to the smallest $N/4$ Fourier modes along
the $\vb{k}_\perp$ axes. With these observational effects applied, we inverse
Fourier transform, and then select the central slice of this slab. This approach
roughly approximates performing a narrow-band measurement of the interferometer,
though the impact of thermal noise has been neglected. Its impact can be fairly
significant for different interferometer designs \citep{hassan_etal2018}, but
for this initial analysis we ignore its effect.

Having generated a series of 20 redshift snapshots in the above manner (which
results in an array with dimensions of $512 \times 512 \times 20$ pixels), the
result is saved as a single ``image'' to be used for training the CNN and
evaluating its performance. The axis of evolution along redshift is treated as
different ``color channels'' in the CNN architecture. Additionally, the
reionization history is computed, and the midpoint and duration are saved. These
values will serve as the ``labels'' when training the CNN. Altogether, we
generate 1,000 of these images to use for training and testing purposes. In
the analysis that follows, we keep $\bar{z}$ fixed, and allow the other
reionization model parameters $\alpha$ and $k_b$ to vary, effectively changing
the duration of reionization while keeping the midpoint nearly constant. This
type of analysis could apply if the midpoint of reionization were derived from,
\textit{e.g.}, power spectrum analysis \citep{lidz_etal2006}, but with the
duration being relatively unconstrained. In future analysis, we plan to extend
this work to regress on the duration and midpoint simultaneously, which would
help serve as an independent confirmation of the reionization history derived
from other methods.

\begin{figure}
  \centering
  \includegraphics[width=0.45\textwidth]{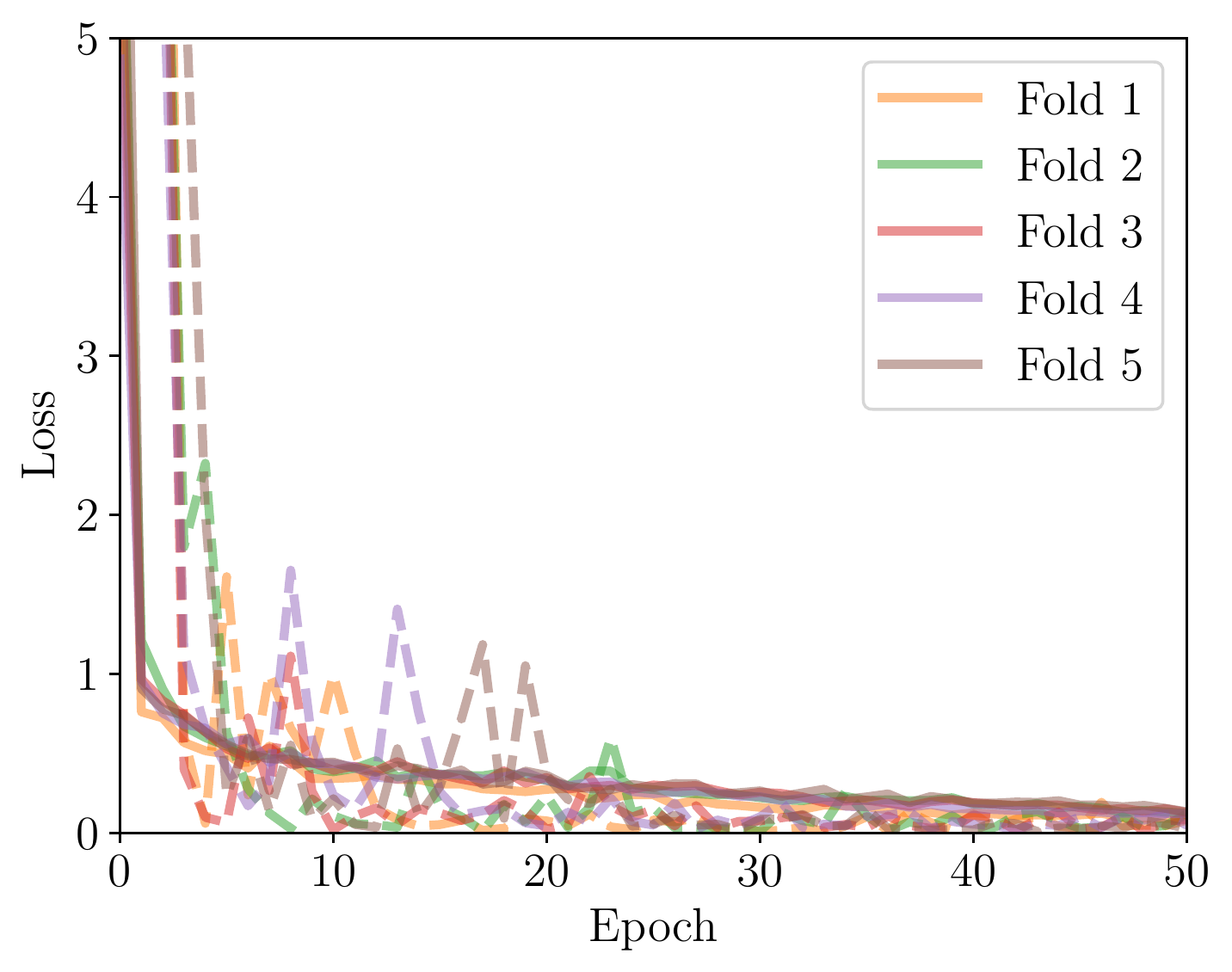}
  \caption{The loss of the CNN network as a function of epoch for the training
    (solid) and test (dashed) sets. The loss is generally decreasing, showing
    that the network is converging on a solution. The loss for the train set is
    not significantly lower than that of the test set, indicating that the model
    is not overfitting. The loss function is presented only out to epoch 50,
    with the full data for all epochs shown in
    Figure~\ref{fig:loss_all_epochs}.}
  \label{fig:loss}
\end{figure}

Figure~\ref{fig:viz_layers} shows a side-by-side comparison of a typical
``image'' (20 redshift layers which serve as a single input image to the CNN),
both with and without the instrumental effects discussed in
Sec.~\ref{sec:foregrounds}. Notably, the zero-level of the image is adjusted
when the foreground effects are applied and the $k_\perp = 0$ modes are
removed. In essence, the bottom row of Figure~\ref{fig:viz_layers} shows just
the dynamic range of the images, rather than an absolutely referenced scale of
the signal, as seen in the top row. In other words, although the reionization
history is encoded in the sky-averaged global signal of the redshifted 21 cm
line, this information is not presented to the CNN trained as part of the
analysis---the CNN only ever receives images like the bottom row as
``input''. Also, the features in the images shift once the foreground effects
are applied, due to the cuts made in Fourier space. Also note that the features
generally become blurrier and less well resolved, and effect of the finite
resolution of HERA.

\section{Machine Learning Techniques}
\label{sec:ml_methods}
Convolutional Neural Networks \citep[CNNs,
e.g.~][]{fukushima_miyake1982,lecun_etal1999,krizhevsky_etal2012}
are a class of feed-forward machine learning algorithms commonly used in image
recognition tasks.  They employ a deep network with hidden layers, and require
very little preprocessing of the input images because the network learns the
convolutional filters necessary to extract relevant features for classification
or regression tasks.

The architecture of the our CNN is shown in Figure \ref{fig:architecture} and is
based loosely on the architecture of \citet{simonyan_zisserman2014}, but with
fewer hidden layers.  Feature extraction is performed with three convolutional
and pooling layers.  The convolutional layers use $3\times3$ convolutional
filters, and are coupled with $2\times2$ max pooling layers with a stride of 2
\citep[e.g.][]{riesenhuber_poggio1999}.  In between the convolutional and max
pooling layers, a batch normalization layer is applied to provide regularization
\citep{ioffe_szegedy2015}. Following feature extraction, regression is performed
by three hidden, fully connected layers.  The fully connected layers employ
rectified linear unit \citep[ReLU,][]{nair_hinton2010} activation.  A 20\%
dropout after each fully connected layer is used to prevent overfitting
\citep{srivastava_etal2014}.

\begin{figure}
  \centering
  \includegraphics[width=0.45\textwidth]{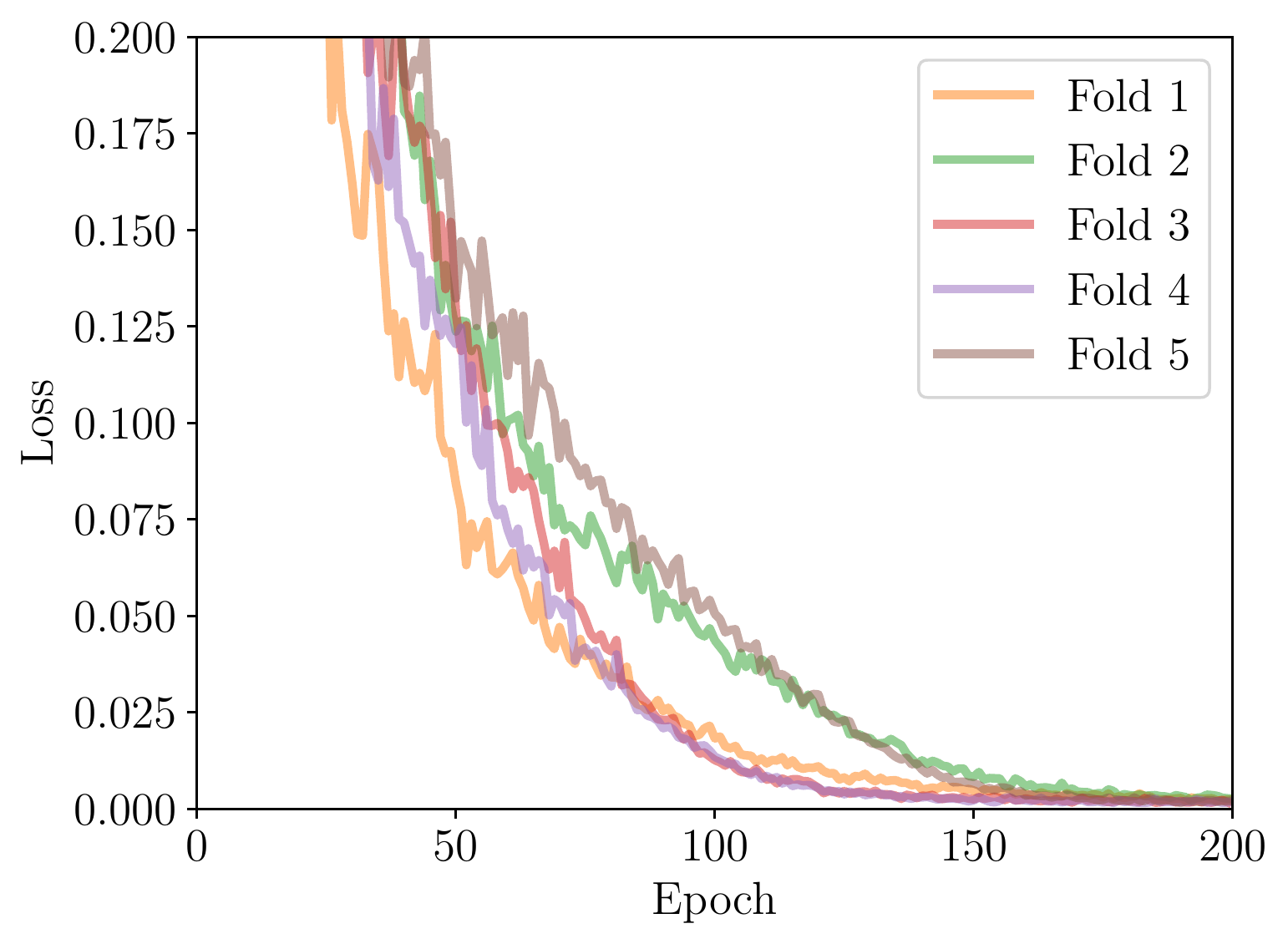}\hspace{10pt}
  \caption{The loss of the CNN network as a function of epoch for the training
    sets for all epochs. The fact that the loss is generally decreasing for all
    epochs shows that the network is converging on a solution. Several of the
    folds converge at earlier epochs, indicating that for certain models the
    full 200 epochs may not have been required.}
  \label{fig:loss_all_epochs}
\end{figure}

The full architecture is as follows:
\begin{enumerate}
\itemsep-0.4em 
\item $3\times 3$ convolutional layer with 16  filters
\item batch normalization
\item $2\times 2$  max pooling layer 
\item $3\times 3$ convolutional layer with 32  filters
\item batch normalization
\item $2\times 2$  max pooling layer 
\item $3 \times 3$ convolutional layer with 64 filters
\item batch normalization
\item $2 \times 2$ max pooling layer
\item global average pooling layer
\item 20\% dropout
\item 200 neuron fully connected layer
\item 20\% dropout
\item 100 neuron fully connected layer
\item 20\% dropout
\item 20 neuron fully connected layer
\item output neurons
\end{enumerate}
See also \cite{schmidhuber2014} and references therein for a review of deep
neural networks.

\begin{figure}
  \includegraphics[width=0.45\textwidth]{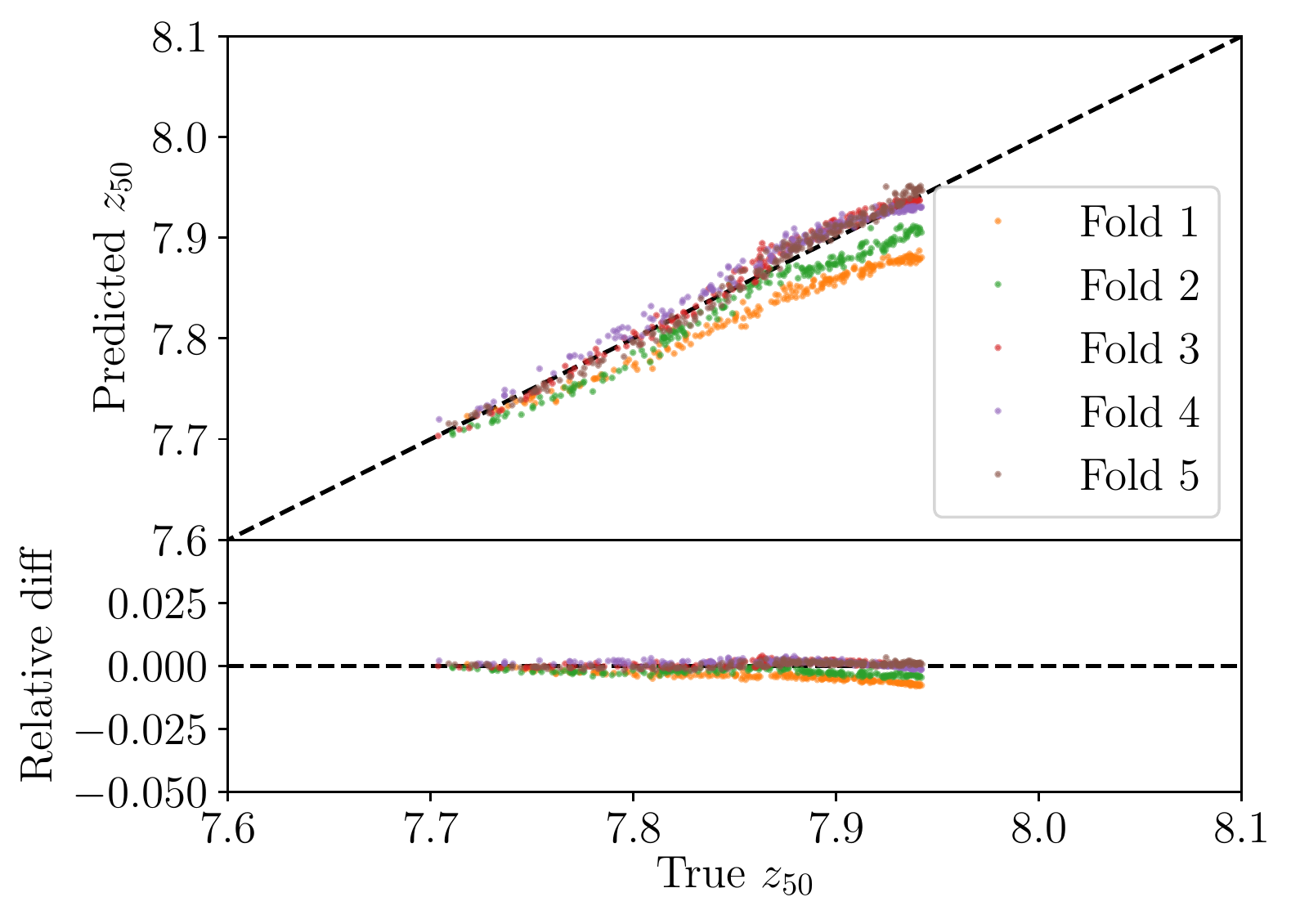}
  \caption{The predicted value of $z_{50}$ as a function of the input (true)
    value of $z_{50}$. These points show that for most of the folds, recovery of
    the input parameter is highly accurate. However, Folds 1 and 2 are biased
    low for large values of $z_{50}$, in that the predicted value for the
    midpoint is systematically lower than the true values. This discrepancy is
    likely due to the relatively large gaps in redshift between layers in the
    training data, which can make precise prediction of the midpoint difficult.
    These errors are typically much smaller than 1\%, even for the biased
    folds. See Sec.~\ref{sec:results} for more discussion.}
  \label{fig:accuracy_zbar}
\end{figure}

We perform a 5-fold crossvalidation, cyclically training on 80\% of the images,
and reporting the results of the remaining 20\%.  Each model is trained for 200
epochs and trains to minimize a root mean squared error (RMSE) loss
function. The CNN architecture described above is implemented using the Keras
package \citep{chollet2015keras}, with the TensorFlow package
\citep{abadi_etal2016} providing the backend computation engine to take
advantage of GPU processing. Also, by construction all images in the training
set had the same mean redshift of reionization $\bar{z} = 8$; however, as
discussed above in Sec.~\ref{sec:zreion}, the midpoint of reionization $z_{50}$
was not identically the same as $\bar{z}$, and varied from
$7.7 \lesssim z_{50} \lesssim 7.9$. Empirically, better results were achieved by
training the CNN to regress on both the midpoint and duration, and we present
results for both parameters below.

Figures~\ref{fig:loss} and \ref{fig:loss_all_epochs} show the loss as a function
of epoch. The solid lines show the loss for the training set and the dashed
lines are the loss for the test set. The oscillations in the test set decrease
significantly after the first $\sim$25 epochs, showing that the majority of the
``learning'' happens during this initial set of epochs. Afterwards, the small
disparity between the training and test set loss values demonstrates that the
CNN is not overfitting the training set. The fact that the loss is generally
decreasing for all epochs shows that the network is converging on a
solution. Although some of the folds seem to have converged before 200 epochs,
the loss does not increase, meaning the networks have not overfitted the data.

\newpage

\section{Results}
\label{sec:results}

\begin{figure}
  \includegraphics[width=0.45\textwidth]{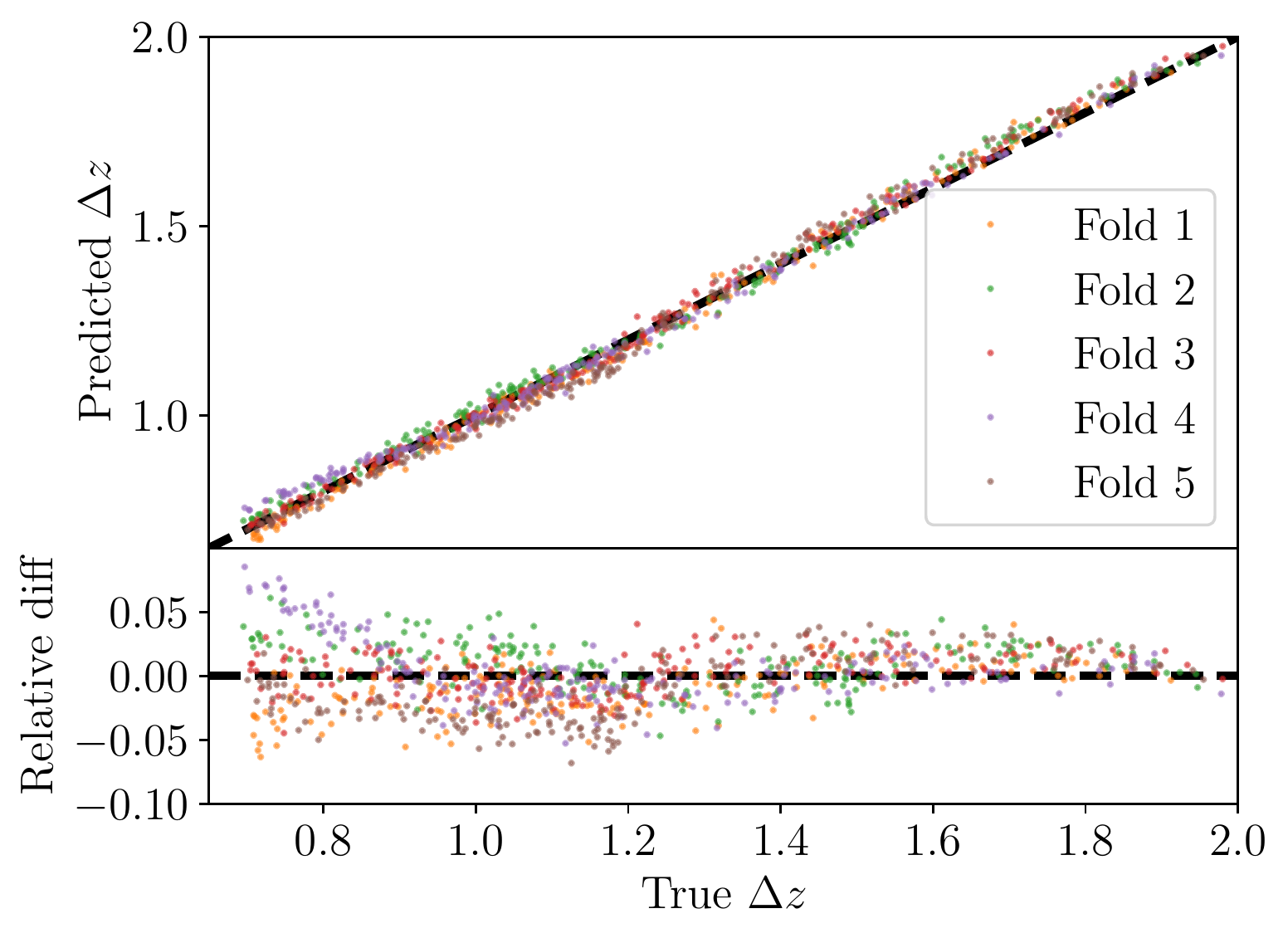}
  \caption{The accuracy of the predicted value of $\Delta z$ as a function of
    the input (true) value of $\Delta z$. The accuracy of the prediction is
    generally within $\pm$5\% of the true value, demonstrating that the CNN can
    accurately predict the duration from the series of training images it
    receives. See Sec.~\ref{sec:results} for more discussion.}
  \label{fig:accuracy_delz}
\end{figure}

As discussed in Sec.~\ref{sec:ml_methods}, the CNNs were trained by minimizing
the RMSE of the predicted reionization parameters relative to the true
ones. Here we present the results of predicting the midpoint $z_{50}$ and
duration of reionization $\Delta z$. Note that both the midpoint and duration
were used as output ``prediction'' neurons, and were regressed on as part of the
RMSE loss function.

Figure~\ref{fig:accuracy_zbar} shows the output of the CNN regression for the
midpoint of reionization $z_{50}$. As discussed in Sec.~\ref{sec:ml_methods},
the CNN produces estimates of both $z_{50}$ and $\Delta z$ for the input images,
despite the fact that $z_{50}$ remained relatively constrained for the different
input images. For most of the folds, the trained network is able to recover the
input value of $z_{50}$ highly accurately, with errors much less than 1\%
percent. However, for Folds 1 and 2, the output values are biased low for high
values of $z_{50}$. This is likely due to the relatively large gaps in between
the central redshifts of the input layers: the gap in redshift between adjacent
layers $\delta z$ is typically $\delta z \sim 0.3$, which can make very precise
prediction of the midpoint difficult. Decreasing the spacing between adjacent
redshift snapshots should help eliminate the bias, though the accuracy of the
current results is still quite good.

Figure~\ref{fig:accuracy_delz} shows the output of the CNN regression for the
duration of reionization $\Delta z$. The top panel of
Figure~\ref{fig:accuracy_delz} shows the absolute deviation from the correct
value, and the lower panel shows the relative deviation of the value of
$\Delta z$. As can be seen in the lower panel, the variation is typically
$\pm$5\% or smaller, implying that the CNN trained using the input data can
predict the duration of reionization reasonably well. As with some of the folds
in Figure~\ref{fig:accuracy_zbar}, some of the folds show a biased result that
changes as a function of the true value of $\Delta z$. Interestingly, this bias
tends not to be consistently positive or negative for all values of $\Delta
z$. As with the bias in Figure~\ref{fig:accuracy_zbar}, the bias is likely due
to the relatively large spaces in redshift space between adjacent
snapshots. Decreasing this spacing may increase the accuracy of the ultimate
predictions.

\begin{figure*}
  \centering
  \includegraphics[width=\textwidth]{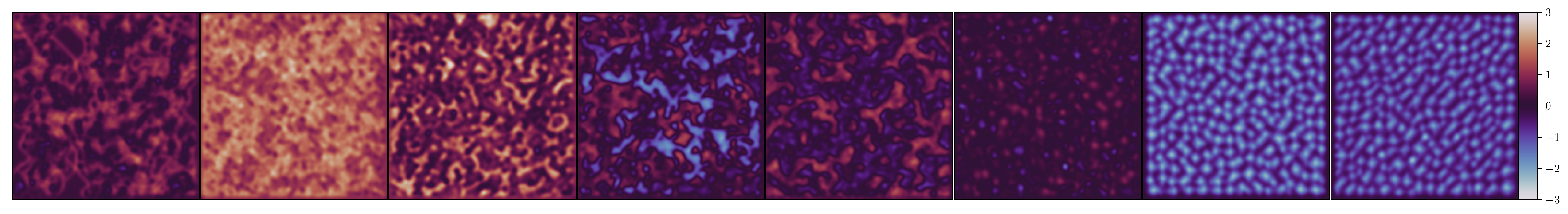}
  \caption{An visualization of the input-layer-maximization approach described
    in Sec.~\ref{sec:discussion}. Each of the two-dimensional images are
    512$\times$512 input layers at the same redshift snapshot ($z \sim 5.8$),
    and are the patterns that maximize the response of different neurons in the
    global average pooling layer (Step~10 in the architecture in
    Sec.~\ref{sec:ml_methods}). Of the 64 output neurons from this layer, 8 of
    the ``typical'' output patterns are presented. The different filters seem to
    be keying in on different features in the images: the patters on the left
    describe large-scale differences in the patterns of the input images,
    whereas the ones on the right identify small-scale contrast. See
    Sec.~\ref{sec:discussion} for additional comments.}
  \label{fig:filters}
\end{figure*}

The results in Figure~\ref{fig:accuracy_delz} show that the CNN is able to
accurately predict the duration of reionization from a single 512 $\times$ 512
$\times$ 20 input image to the CNN. Given the field-of-view of HERA and its
observation strategy, it is expected to generate a modest number (10-20) of such
independent observations. Assuming that statistical errors dominate the
uncertainty over systematic errors, the ultimate accuracy with which HERA would
be able to constrain the duration of reionization $\Delta z$ should be even
better than the 5\% show in the figure. Part of this extrapolation assumes that
the duration of reionization is well-characterized by the semi-analytic model
used in this analysis, and so in future work we plan to use additional
semi-analytic tools, such as 21\textsc{cmfast} \citep{mesinger_etal2011} to test
whether the results are robust when using a different semi-analytic model. As
discussed in Sec.~\ref{sec:foregrounds}, we also plan to incorporate the effect
of thermal noise from interferometer measurements in future analysis.

\section{Discussion}
\label{sec:discussion}
Gaining physical insight from deep networks is difficult, but not impossible.
CNNs are often utilized as a black box that takes an unprocessed image as an
input and outputs an image label, but this need not be the case.  Filter
visualization is one way to interpret these systems to add a human layer of
interpretability and physical understanding to the tool.

Motivated by the work in \citet{cnn_see} to visualize filters of an image
classifier, we explore the types of input images that maximize neurons of the
global average pooling layer. This type of analysis is useful for understanding
the types of images that the CNN has been trained to interpret, and can be
useful for seeing which types of features are most important to the CNN. Other
approaches to visualizing ``what the CNN sees'', such as passing the input
images through one or several convolutional layers, can be useful for
understanding how the images are modified as they pass through the series of
convolutional filters. However, they do not necessarily show which features in
particular the network is responding to. The reason for this is that those types
of analysis rely on permuting some typical input, rather than allowing the input
images to maximize the response of a given neuron deep in the network. Though
the visualization approaches are related, they are fundamentally showing
different aspects of the CNN learning process.

We create an input image that is the same size ($512\times512\times20$) as one
training image.  These images are initially white noise; for each of the 20
redshift slices, the pixels are populated by random numbers that span a typical
pixel value range of the input data at that redshift.  Through the iterative
process described in detail in \citet{cnn_see}, the pixel values of these noisy
input images are altered through gradient ascent over 10 iterations to maximize
the response at the global max pooling layer.

After each iteration, we employ a normalization to prevent single pixels from
dominating the effect as well as a smoothing to make the resulting images easier
to visually interpret.  The normalization is achieved simply by multiplying each
pixel value by $0.8$ to minimize very bright pixels.  The smoothing is
accomplished with a Gaussian blur with $\sigma=5$ pixels that extends
along the $512\times512$ single-redshift image but does not blur one redshift
slice into another.

A sample of these images is shown in Figure \ref{fig:filters}. This
visualization shows the input images that maximize the response to 8 different
nodes deep in the CNN, for the same redshift layer (one corresponding to
$z \sim 5.8$). Interestingly, the shapes and patterns for a given node do not
vary significantly across the different redshift layers. This behavior suggests
that each neuron is responding to a particular pattern in the input data at
different redshifts, though the relative importance between these patterns can
change by changing the weights. We have selected several of the typical patterns
seen for the different nodes, which represent nearly all of the different ones
seen in the various filters. We may be able to ascribe some physical
interpretation to the different patterns, which will help in understanding which
trends are most important to the CNN when regressing on the duration of
reionization.

The first few filters on the left side of the figure show, broadly, variations
in the large-scale structure present in the input maps. The input images to the
CNN represent comoving volumes that are 2~$h^{-1}$Gpc in size, and so these
large scale fluctuations are tens to hundreds of Mpc on a side. Rather than
detecting individual ionized regions, these filters seem to be picking up on the
large-scale contrast present in the maps. The size of the connected regions can
also change, and seem to range from large scales, to intermediate scales, and
finally relatively small scales. Additionally, the dynamic range and sign of the
contrast can change for the different filters, suggesting that different neurons
are not equally sensitive to all types of contrast.

In the two rightmost panels of Figure~\ref{fig:filters}, there are two input
images that focus on the small-scale contrast of the input images. However,
there is a distinct difference in the shape of this small-scale structure: one
has more circular features, and the other has elliptical ones. This may be a
reflection of the CNN using the degree of anisotropy of the small-scale
structure as a way of determining the duration of reionization. As seen in
Figure~4 of \citet{laplante_etal2014}, shorter reionization scenarios tended to
have more anisotropic features compared to longer reionization histories. Though
this may be a feature of the common semi-analytic model used in both works, it
nevertheless points to an interesting feature that the CNN may be using to
determine the reionization history. More generally, the degree of anisotropy in
the 21 cm maps may hint at the relative bias of the luminous sources
contributing to reionization. If these types of features are present in similar
analysis employed by other semi-analytic models of reionization, it may point to
a common underlying physical property being used by the CNN. We plan to pursue
such comparisons in future work.

\section{Conclusion}
\label{sec:conclusion}
In this work, we show that we are able to train a CNN to accurately predict the
duration of the EoR to 5\% or better. The input images for this task have been
modified to reflect some of the effects that foreground avoidance and removal
strategies in HERA data processing are expected to impart to real-world images
generated in the next few years, though the inclusion of thermal noise is saved
for future work. We show that despite the degradation in input image quality,
the CNN is able to extract the target parameter with reasonable accuracy. The
ultimate accuracy of such a method will be improved by combining independent
image cubes generated from statistically independent portions of the
sky. Constraining the duration of reionization with a relatively high degree of
accuracy has exciting implications on the ability to provide priors on $\tau$,
the optical depth to the CMB. Such a constraint can be used to decrease the
uncertainty of other cosmological parameters, such as $A_S$, which are partially
degenerate with $\tau$.

\acknowledgments{We thank James Aguirre, Daniel Eisenstein, Anastasia Fialkov,
  Josh Kerrigan, Francois Lanusse, Avi Loeb, Junier Oliva, Hy Trac, and
  Kun-Hsing Yu for their helpful feedback on this project. This material is
  based upon work supported by the National Science Foundation under Grant
  No. 1636646, the Gordon and Betty Moore Foundation, and institutional support
  from the HERA collaboration partners. HERA is hosted by the South African
  Radio Astronomy Observatory, which is a facility of the National Research
  Foundation, an agency of the Department of Science and Technology. This work
  was supported by the Extreme Science and Engineering Discovery Environment
  (XSEDE), which is supported by National Science Foundation grant number
  ACI-1548562 \citep{xsede2014}. Specifically, it made use of the Bridges
  system, which is supported by NSF award number ACI-1445606, at the Pittsburgh
  Supercomputing Center \citep{bridges2015}.}

\bibliography{mybib}
\bibliographystyle{apj}

\end{document}